\documentclass[traditabstract]{aa}
\usepackage{epsfig}
\usepackage{rotating}
\usepackage{natbib}
\usepackage{graphicx}
\bibpunct{(}{)}{;}{a}{}{,} 

\newcommand{\chan}{{\sl Chandra}}

\newcommand{\hstn}{{\sl Hubble Space Telescope}}
\newcommand{\hst}{{\sl HST}}
\newcommand{\vltn}{{\sl Very Large Telescope}}
\newcommand{\vlt}{{\sl VLT}}
\newcommand{\wfpcn}{{\sl Wide Field Planetary Camera 2}}
\newcommand{\wfpc}{{\sl WFPC2}}
\newcommand{\acsn}{{\sl Advanced Camera for Surveys}}
\newcommand{\acs}{{\sl ACS}}

\newcommand{\wfc}{{\sl WFC}}
\newcommand{\isaacn}{{\sl Infrared Spectrometer And Array Camera }}
\newcommand{\isaac}{{\sl ISAAC}}

\def \onee{1E\,1207.4$-$5209}

\begin{document}

 \title{HST and VLT observations of  the neutron star 1E\,1207.4$-$5209\thanks{Based on observations with the NASA/ESA \hstn, 
obtained at the Space Telescope Science Institute, which is operated by
AURA,Inc.\ under contract No NAS 5-26555; Based on observations collected at ESO, Paranal, under Programme 70.D-0436(A)}}

 \author{
 A. De Luca\inst{1,2,3}
  \and 
 R.~P. Mignani\inst{4}
 \and
 A. Sartori\inst{1,5}
 \and
W. Hummel\inst{6}
\and
P.~A. Caraveo\inst{2}
\and
S. Mereghetti \inst{2}
\and
G.~F. Bignami \inst{1}
}

\institute{
Istituto Universitario di Studi Superiori, Viale Lungo Ticino Sforza 56, I-27100 Pavia, Italy \\
\email{deluca@lambrate.inaf.it}
\and
INAF, Istituto di Astrofisica Spaziale e Fisica Cosmica di Milano, Via Bassini 15, I-20133 Milano, Italy 
\and
Istituto Nazionale di Fisica Nucleare, Sez. di Pavia, Via Bassi 6, I-27100 Pavia, Italy 
\and
Mullard Space Science Laboratory, University College London, Holmbury St. Mary, Dorking, Surrey, RH5 6NT, UK
\and
Universit\`a degli Studi di Pavia, Dipartimento di Fisica Nucleare e Teorica,
Via Bassi 6, I-27100 Pavia, Italy 
\and
European Southern Observatory, Karl  Schwarzschild-Str. 2, D--85748, Garching, Germany
}

\titlerunning{HST and VLT observations of \onee}
\authorrunning{A. De Luca et al.}
\offprints{deluca@lambrate.inaf.it}

     \date{Received ...; accepted ...}


\abstract{  \onee,   the  peculiar  Central  Compact   object  in  the
G296.5+10.0   supernova  remnant,   has   been  proposed   to  be   an
``anti-magnetar''  -- a  young neutron  star born  with a  weak dipole
field. Accretion,  possibly of  supernova fallback material,  has also
been invoked to explain a large surface temperature anisotropy as well
as   the  generation   of  peculiar   cyclotron   absorption  features
superimposed to  its thermal  spectrum. Interestingly enough,  a faint
optical/infrared  source was  proposed  as a  possible counterpart  to
\onee, but  later questioned, based on  coarse positional coincidence.
Considering the large offset of \onee\ with respect to the center
of its host supernova remnant, the source should move at $\sim70$
mas~yr$^{-1}$. Thus, we tested the association by measuring the proper motion
of the proposed optical counterpart.
Using \hstn\  (\hst) observations spanning  3.75 years, 
we computed a $3\sigma$ upper limit of 7 mas~yr$^{-1}$.
Absolute astrometry on  the same
\hst\ data  set also  places the optical  source significantly  off the
99\% confidence {\em Chandra} position.  This allows us to safely rule
out  the   association.   Using  the   \hst\  data  set,   coupled  to
ground-based observations  collected at the ESO/\vltn\  (\vlt), we set
the deepest limits ever obtained to the optical/infrared emission from
\onee.  By combining such limits to the constraints derived from X-ray
timing, we rule out accretion  as the source of the thermal anisotropy
of the neutron star.  }


   \maketitle

\section{Introduction} 

The X-ray source
\onee\ was  discovered with the  {\em Einstein} satellite  (Helfand \&
Becker 1984)  close to  the center of  G296.5+10.0, a $\sim7$  kyr old
supernova  remnant  (SNR)  located   at  a  distance  of  $\sim2$  kpc
\citep{roger88,giacani00}.
It  was the second  thermally-emitting, radio-quiet,  Isolated Neutron
Star (INS) candidate  found inside a SNR, after  1E\, 161348$-$5055 in
RCW  103 \citep{tuohy80}.  A  handful of  similar sources,  discovered
inside young  SNRs, are  dubbed, as a  class, Central  Compact Objects
\citep[CCOs,][for a recent review]{pavlov02,deluca08a}.  Although their
properties  are not  well  understood,  CCOs are  supposed  to be  the
youngest members of the radio-quiet INSs family.

\onee\ is  one of the most  peculiar and most  observed Galactic X-ray
sources. Pulsations at  424 ms were discovered with  the {\em Chandra}
satellite \citep{zavlin00},  proving the source  to be an  INS.  Early
timing investigations  hinted at a non-monotonic  period evolution of
\onee\ suggesting  that the source  could be a peculiar  binary system
\citep{zavlin04,woods07}.  However,  more recently, \citet{gotthelf07}
provided robust evidence that \onee\ is a very stable rotator.
The upper limit  to its period derivative ($\dot{P}<2.5\times10^{-16}$
s s$^{-1}$ at 2$\sigma$)  yields an INS carachteristic age $\tau_c>27$
Myr, exceeding by 3 orders of magnitude the age of the SNR, a rotational 
energy loss $\dot{E} < 1.3 \times10^{32}$ erg s$^{-1}$,  and a very
small  dipole  magnetic field,  B$<3.3\times10^{11}$  G. Thus,  \onee\
could be a young, weakly magnetized INS, born with a spin period very similar
to the current one.
Evidence for  similar, low  magnetic fields has  been obtained  for two
other members of the CCO  class, namely CXOU\, J185238.6+004020 at the
center of  the Kes 79  SNR \citep{halpern07} and RX\,  J0822$-$4300 in
Puppis A \citep{gotthelf09}, adding support to the scenario of CCOs as
``anti-magnetars''  \citep[with   the  remarkable  exception   of  the
puzzling source in RCW 103,][]{deluca06}.

What makes \onee\ unique among all INSs is its 
X-ray spectrum. Three (possibly  four) broad absorption features -- at
regularly spaced energies (0.7, 1.4,  2.1 and possibly 2.8 keV) -- are
visible      over     a     thermal      two-temperature     continuum
\citep{mereghetti02,sanwal02,bignami03,deluca04}.  The  depth of  such
features   varies   as   a    function   of   the   rotational   phase
\citep{mereghetti02,deluca04}.
The nature of  the spectral features of \onee\  has been debated since
their  discovery,  possible  interpretations being  atomic  transition
lines  in  the NS  atmosphere  or  cyclotron  features in  the  plasma
surrounding   the  star   \citep{sanwal02,mereghetti02}.   The  latter
interpretation,  favoured  by  the  harmonic  energy  spacing  of  the
features  \citep{bignami03,deluca04},  is  fully consistent  with  the
emerging  picture  of \onee\  as  a  weakly  magnetized neutron  star:
assuming the 0.7 keV feature  to be the fundamental electron cyclotron
line yields  a measure  of the magnetic  field of  $8\times10^{10}$ G,
i.e. below the  value derived from the upper limit  on the pulsar spin
down.

Many  puzzles   remain  to  be   solved  for  \onee.   As   stated  by
\citet{gotthelf07},  it is  difficult  to explain  the luminosity  and
temperature of the observed  hot thermal spectral component within the
frame of a weakly magnetized INS.
Moreover,  the  physics underlying  the  peculiar absorption  features
\citep[most          likely          due         to          cyclotron
processes,][]{liu06,suleimanov10,potekhin10} has to be understood.  The
simplified  model  proposed  by  \citet{liu06} requires  a  high  (and
steady) electron density in the  NS magnetosphere above the polar cap.
Low-level accretion, possibly of supernova fallback material, has been
invoked to  ease the problem  in both cases.  This would point  to the
existence  of  a  debris  disk  surrounding the  INS (which could
be detected in the optical-infrared range), a  long  sought
astrophysical object, so far possibly observed only in the case of the
Anomalous X-ray Pulsar (AXP)  4U\, 0142+61 \citep{wang06}. 

Very deep imaging of the field of \onee\ have been performed both from
the ground with  the ESO \vltn\ (\vlt) and with  the {\em Hubble Space
Telescope} (\hst).  Optical \vlt\ observations
(De Luca et al. 2004) did not reveal any potential counterpart down to
R$\sim$27.1 and V$\sim$27.3
while observations in the optical with the \hst\
and in the near infrared (NIR) with the \vlt\
showed the presence of a faint source (hereafter ``source Z'')
close to the
{\em Chandra} X-ray position, with magnitudes m$_{F555W}\sim26.4$
and  Ks$\sim$20.7 \citep{pavlov04,fesen06}.  However,  the association
with  \onee\ was soon  after questioned  by \citet{mignani07a}  on the
basis of precise absolute astrometry  of the \hst\ images, which showed
a positional offset  of source Z  with respect to  the {\em
Chandra}  coordinates. The  same source  was  observed in  the NIR  by
\citet{wang07}, who  reported very red  colours, consistent with  an M
dwarf, and also questioned  its possible association to \onee\ because
of
the  inconsistency with  the {\em  Chandra}  position.  \citet{wang07}
also  observed  the field  with  {\em  Spitzer}  at 4.5$\mu$m  and  at
8.0$\mu$m, but did not detect any source at the target position.

Here  we  report  on  a  different, independent  test  to  assess  the
association  of source Z  to \onee,  using multi-epoch  data collected
with the \hst\ (Sect.~\ref{hst}). The same \hst\ dataset, completed by
ground  based data collected  with the  \vlt, is  also used  to derive
stringent  constraints on  the optical/infrared  emission  from \onee\
(Sect.~\ref{phot}). Results are discussed in Sect.~\ref{disc}.


\section{Association of source Z to \onee: an HST test.}

\label{hst}
G296.5+10.0  has a
remarkable,        well        defined       bilateral        symmetry
\citep{roger88,storey92}. Very likely, the  explosion site lies on the
symmetry axis of the SNR, but the current  position of \onee\  is significantly
offset from  the apparent  center of the  host SNR. 
Indeed, the geometrical center position evaluated by
\citet{roger88}  is $\sim8\arcmin$  to  the south  west  of the  X-ray
source  \citep{deluca04}. Assuming  for the  system an  age  of 7\,000
years, such  a displacement  would imply a  proper motion  of $\sim70$
mas~yr$^{-1}$,  corresponding  to a  projected  velocity of  $\sim640$
km~s$^{-1}$,  consistent with the  observed velocity  distribution for
radio  pulsars \citep{hobbs05}.  This  offers  a  natural way  to  test  the  putative
identification: if indeed associated  with \onee, source Z should show
a  significant   proper  motion.  Thus,  we   used  multi-epoch  \hst\
observations to search for an angular displacement of source Z.

\subsection{\hst\ observations}

We  observed the  field  of \onee\  with  the \hst\  on  May 8th  2007
(Programme 10791).  Our  observations were originally scheduled for
execution with the {\em Wide  Field Channel} ({\em WFC}) of the \acsn\
(\acs) \citep{clampin00,sirianni05}.
Unfortunately, the {\em ACS/WFC} was put in idle state on January 2007
due to a failure of  the on-board electronics. 
Our  observations  were re-scheduled  and  executed  with the  \wfpcn\
(\wfpc).   A set  of  four  500s exposures  were  obtained during  one
spacecraft orbit, through the 814W filter ($\lambda= 8012$\AA; $\Delta
\lambda=1539$ \AA).
In  order to  exploit the  maximum spatial  resolution for  the proper
motion  measurement, \onee\  was  placed  at the  centre  of the  {\em
Planetary Camera} ({\em PC}) chip (0\farcs045/pixel).

Our new data  add up to observations collected with  the \acs\ on July
28th  and  August  7th  2003   (Programme 9872) and available in the public 
\hst\ archives.  
This first-epoch \hst\ dataset allowed \citet{pavlov04} 
to pick up Source Z as a possible counterpart to \onee.
The  {\em  WFC}
(0\farcs050/pixel) was used in both visits.
Two  sequences  of  4  and  5  exposures  were  obtained  through  the
broad-band  filters 555W  ($\lambda=  5346$\AA; $\Delta  \lambda=1193$
\AA) and  814W ($\lambda= 8333$\AA; $\Delta \lambda=2511$  \AA), for a
total  integration time  of 12800  s and  10200 s,  respectively.  The
complete dataset spans a time baseline of $\sim3.75$ years.

We downloaded the data  from the Space Telescope European Coordinating
Facility         (ST-ECF)               Science         Data
Archive\footnote{www.stecf.org/archive}.   On-the-fly  data  reduction
(bias  and flat-field  correction) and  flux calibration  were applied
using the  {\em Space Telescope Science Data  Analysis Software} ({\em
STSDAS}) through  the ST-ECF Data Archive pipeline.   To filter cosmic
ray hits, single \wfpc\ exposures were combined and averaged using the
{\em STSDAS}  task {\tt combine}, while single \acs\  exposures were
combined using  {\tt multidrizzle} which also produces  a mosaic image
of the  two \acs\ chips and  applies the correction  for the geometric
distortions of the camera.

\begin{figure}
\centering 
\includegraphics[width=9.0cm,clip]{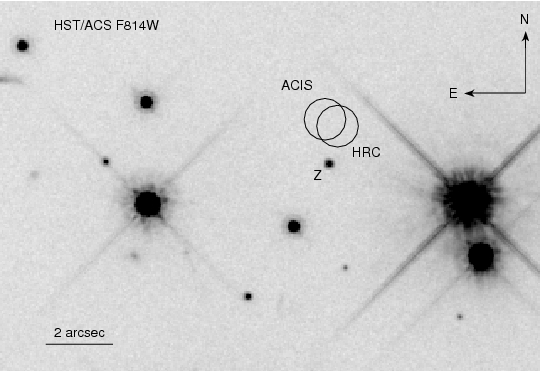}
\caption{$16\arcsec \times 11\arcsec$ cutout of the 2003 \acs\ image of  \onee\  obtained through the 814W filter. The {\em Chandra} position of the target, as computed using the {\em ACIS} and the {\em HRC} observations, is shown. The circles (0\farcs62 radius)  corresponds to the uncertainty of the CCO position computed from the error  of the \chan\ coordinates and the accuracy of the 2MASS absolute astrometry.  North to the top, East to the left.  }
\label{acs}       
\end{figure}

\subsection{Relative astrometry}

We have used the algorithm that we successfully applied in a series of
previous   proper   motion    measurements   with   \hst\   \citep[see
e.g.][]{deluca99,caraveo01,mignani02,deluca07,mignani10}. We used only
the  \acs\  F814W  bandpass   image  in  order  to  avoid  introducing
systematic effects in the comparison of images taken through different
filters. Briefly, a  relative reference frame is defined  by a grid of
good (not extended,  with good signal to noise  but not saturated, not
too close to the CCD edges) reference sources common to all images and
spread homogeneously across the common  field of view.  We selected 32
such sources and we computed their positions by fitting a 2-D gaussian
profile to their brightness profile. The centering accuracy was between
0.02 and 0.07 pixels per coordinate for the {\em WFC} and between 0.04
and 0.08 pixels for the {\em PC}
 depending  the  source  brightness.  The  position of  source  Z  was
evaluated in the same way, with an uncertainty of 0.08 and 0.04 pixels
in the  {\em PC} and in  the {\em WFC} image  respectively.  Next, for
the {\em PC} image, we corrected the pixel coordinates for the effects
of  the ``34th  row effect''  \citep{anderson99}  as well  as for  the
geometric  distortion \citep{anderson03}.  Then, we  assumed  the 2007
image as  a reference and we  aligned the coordinate  grid along Right
Ascension  and Declination  using the  well-measured \hst\  roll angle
($\la 0.1^{\circ}$ uncertainty). We then registered the 2003 reference
frame  to  the  2007  one  by  fitting  a  5-parameter  transformation
(including x  and y shifts  and scale factors,  as well as  a rotation
angle). We  rejected 5 objects yielding $>3\sigma$  residuals using an
iterative   $\sigma$-clipping   algorithm.   We   ended  up   with   a
r.m.s.  residual of  0.17  {\em   PC}  pixels  on the frame registration 
 using   27  reference
sources. Finally, we applied  the transformation to the coordinates of
source Z  in order to  evaluate its possible displacement  between the
two epochs. However, such a  displacement was found to be much smaller
than  the overall  uncertainty of  0.19 {\em  PC} pixels.  Taking into
account the {\em PC} plate scale ($0\farcs04554$/pixel) as well as the
time baseline  covered by  our dataset, we  can set a  $3\sigma$ upper
limit of  7 mas~yr$^{-1}$  to the overall  proper motion of  source Z.
This limit is much lower than  the expected proper motion of $\sim$ 70
mas~yr$^{-1}$ implied by the association between \onee\ and the SNR.

We  note that the  proper motion  test relies  on assumptions  for the
supernova remnant age and expansion  center position. The age is quite
uncertain.  However,  even  assuming  a  value  3  times  higher  than
estimated  by \citet{roger88},  we  would expect  a  proper motion  of
$\sim25$ mas~yr$^{-1}$. 

The  good bilateral  symmetry of G296.5+10.0 suggests that the true  
expansion center should lie  not far 
off the apparent symmetry axis. 
However, expansion of the supernova 
remnant  could have  been  anisotropic. 
Indeed, \citet{roger88} report a smaller radius of curvature
for the Eastern side of the SNR with respect to the Western side, 
suggesting an expansion center to the East of the geometrical center 
(i.e. closer to the CCO than the geometrical center), 
although they did not quantify such effect. 
We might assume that
the true expansion center is placed halfway between \onee\ and the SNR
geometrical center, yielding a factor 2 error in our estimate. Even in
this case  (and even combining  the two pessimistic  assumptions), the
expected  proper  motion  would   have  been  larger  than  our  upper
limit. Thus, this  strongly argues against association of  source Z to
\onee.

\subsection{Absolute astrometry}

A  second,  model-independent, test  on the  possible association
between \onee\ and source Z is obviously represented by the comparison
of  their positions.  Before  registering the  \onee\ position  on our
\hst\ images we  decided to recompute anew its  X-ray coordinates from
the {\em  Chandra} data, to account for  possible corrections in
the satellite aspect  solution.  To this  aim, we retrieved  all the archival
{\em Chandra} imaging observations of \onee. The first observation was
collected  with  the {\em  Advanced  CCD  imaging Spectrometer}  ({\em
ACIS-S}) on  2003, June 6  (Obs.Id. 3913), totalling $\sim20$  ks; the
second one was  collected with the {\em High  Resolution Camera} ({\em
HRC-I})  on  2003, December  28  (Obs.Id.  4593)  and lasted  $\sim50$
ks. Data analysis  was performed as in \cite{deluca09}.   We note that
while  Mignani   et  al.  (2007)  used  only  the
coordinates derived from the {\em ACIS-S} data set, Wang et al. (2007)
used  coordinates  derived  from  both  data  sets.   
The position of \onee\  turned out to
be   $\alpha(J2000)$=12$^h$   10$^m$   0.92$^s$,   $\delta   (J2000)$=
-52$^\circ$  26\arcmin 28\farcs40  in the  {\em ACIS}  observation and
$\alpha(J2000)$=12$^h$ 10$^m$  0.88$^s$, $\delta (J2000)$= -52$^\circ$
26\arcmin 28\farcs61 in the  {\em HRC} observations. The two positions
agree  within  $0\farcs4$,  as  expected according  to  {\em  Chandra}
astrometric     accuracy     ($0\farcs6$     at    90\%     confidence
level\footnote{http://cxc.harvard.edu/cal/ASPECT/celmon/})   and   are
consistent with those computed by \citet{wang07} using the same data.
To evaluate the accuracy of  the {\em Chandra} absolute astrometry, we
cross-correlated  the  position   of  X-ray  sources  detected  within
3\arcmin\ of the optical axis with that of stars in the Two Micron All
Sky  Survey \citep[2MASS,][]{skrutskie06}  catalog. We  found two
matches  between 2MASS  stars and  sources detected  both by  the {\em
ACIS} and {\em  HRC}. Based on the source density,  we expect a chance
alignment of {\em  one} 2MASS source to a  {\em Chandra} source within
the nominal $90\%$ error region to  have a probability of 0.6\% and of
0.2\% for  {\em ACIS}  and {\em HRC},  respectively. Thus, it  is very
likely  that the  two 2MASS  sources are  the IR  counterparts  of the
matched  {\em  Chandra}  sources.   The difference  between  the  {\em
Chandra}   and  the  2MASS   coordinates  of   these  sources   is  of
$\sim0\farcs4$  and  is   consistent  with  the  expected  astrometric
accuracy of {\em Chandra}.  Offsets along Right Ascension and 
 Declination range from 
$0\farcs07$ to $0\farcs33$ and have different directions.
Thus, no significant plate transformation could be computed in 
order to further improve Chandra astrometry. Use of the USNO-B1 
catalog \citep{monet03} yields one further possible 
coincidence (with a larger 
offset $\sim1\farcs1$), which is of no help.

On the  optical side, we then re-computed  the astrometric calibration
of the large-field of view \acs\ image against the positions of
35 well-suited  reference stars (i.e.,  not too faint,  not saturated,
not close  to the  CCD edges or  to diffraction spikes)  selected from
2MASS
\footnote{\citet{mignani07a} used  an early release of  the Guide Star
Catalog  2  \citep{lasker08}, while  \citet{wang07}  used the  USNO-A2
Catalog \citep{monet98}.}   identified in the mosaic of  the two \acs\
chips.  We measured the pixel coordinates of the 2MASS sources through
gaussian fitting with the {\em Graphical Astronomy and Image Analysis}
({\em                                                            GAIA})
tool\footnote{star-www.dur.ac.uk/$\sim$pdraper/gaia/gaia.html}  and we
computed  the pixel-to-sky coordinates  transformation using  the code
{\tt ASTROM}\footnote{http://star-www.rl.ac.uk/Software/software.htm}.
This yielded an rms of $\sigma_r \approx 0\farcs11$ in our astrometric
fit, accounting  for the  rms of  the fit in  the right  ascension and
declination components.   Thanks to  the pixel scale  of the  \wfc, we
neglected the uncertainty on  the reference star centroids.  Following
\citet{lattanzi97},   we  also  estimated   the  uncertainty   in  the
registration of the \acs\ image on the 2MASS reference frame.  This is
given as  $\sigma_{tr}=\sqrt 3  \times \sigma_{ref} /  \sqrt N_{ref}$,
where $\sqrt  3$ accounts for  the free parameters in  the astrometric
fit, $\sigma_{ref} \la  0\farcs2$ is the mean positional  error of the
2MASS coordinates, $N_{ref}$ is the number of 2MASS stars used for the
astrometric  calibration.   In   our  case,  we  obtain  $\sigma_{tr}=
0\farcs06$. We  finally considered  the 0\farcs015 uncertainty  on the
link of  2MASS to the International Celestial  Reference Frame (ICRF).
Thus, by adding  in quadrature the rms of the  astrometric fit and all
the  above  uncertainties, we  obtained  that  the overall  positional
accuracy   of  our   \acs\   astrometry  is   $\delta  r   =0\farcs13$
($1\sigma$).  By  finally  adding  in  quadrature this  value  to  the
$0\farcs6$  error on  the \chan\  coordinates we  obtained  an overall
uncertainty  of $\sim  0\farcs62$ on  the registration  of  the \onee\
position on the \acs\ image.

Results are shown in  Figure~\ref{acs}. {\bf Source Z} lies about $1\farcs1$
and  $1\farcs3$  off the  {\em  Chandra}  {\em  HRC} and  {\em  ACIS}
position  of \onee,  respectively. Such  offsets are  larger  than the
expected 99\% accuracy of Chandra astrometry, which is estimated to be
of $0\farcs8$  close to the centre  of the field of  view.  As already
concluded  by   Mignani  et   al.  (2007a)  and   \citet{wang07},  the
association  of source  Z to  \onee\ based  on  positional coincidence
seems very unlikely.

\section{Photometry}

\label{phot}
Results  from our  HST test  allow us  to exclude  any  association of
source Z to \onee.  Thus, we can use the deep {\em  ACS} images to set
upper limits to any undetected source at the position of \onee.
%
%
We focus on the two deep \acs\ observations obtained through the F555W
and  F814W  filters. To  estimate  count  rates,  we used  a  circular
aperture  of   $0\farcs5$  radius  and  then   we  performed  aperture
correction  following  \citet{sirianni05}.   Count rate  to  magnitude
conversions was performed using standard \acs\ photometric calibration
provided  by the data  processing pipeline.   Taking into  account the
observed background  noise in a  portion of the image  surrounding the
position of \onee, we set  a $3\sigma$ upper limit of m$_{F814W}$=28.1
and m$_{F555W}$=28.1.  For completeness, we also computed  the flux of
source   Z,  which   resulted  to   be   m$_{F814W}=24.71\pm0.01$  and
m$_{F555W}=26.80\pm0.05$.  Using  the  more  recent \wfpc\  image,  we
computed m$_{F814W}=24.83\pm0.08$,  consistent with the  \acs\ flux in
the same band.
The quoted values are not dereddened.

Moreover,  we  include  in  our  photometric  study  a  series  of  IR
observations collected with the \vlt\ and available in the public ESO archive
(Programme 70.D-0436A). 
\citet{pavlov04} and \citet{fesen06} reported preliminary results from such data.
%
%
IR  observations of  \onee\ were  performed in  service mode  with the
\isaacn\  (\isaac)  instrument  at  the  \vlt\  (Paranal  Observatory)
between  January  27th  and   March  18th  2003.   The  \isaac\  Short
Wavelenght (SW) camera
was used,  with a projected  pixel size of  0\farcs148 and a  field of
view  of 152$\times$152 arcsec.   Observations were  performed through
the   $J$  ($\lambda=   1.25  \mu$;   $\Delta  \lambda=   0.29  \mu$),
$H$($\lambda=  1.65  \mu$;  $\Delta  \lambda=  0.30  \mu$)  and  $K_s$
($\lambda= 2.16  \mu$; $\Delta \lambda=  0.27 \mu$) band  filters.  To
allow  for  the  subtraction   of  the  variable  IR  sky  background,
observations  in  each filter  were  split  in  sequences of  shorther
dithered exposures with integration times of 50 s in the $H$ and $K_s$
bands and of 120  s in the $J$ band along each  point of the dithering
pattern.  The journal of observations is reported in Table 1.

\begin{table}
\begin{center}
  \caption{Summary of  the \vlt/\isaac\ observations of  the \onee\  field, with the observing epochs, the filter, the total integration time, the average seeing, and airmass  values. }
\begin{tabular}{cccrcc} \\ \hline
yyyy-mm-dd     & Filter & T (s) & Seeing (``) & Airmass	\\ \hline
2003-01-27   &      J     &  1200    &   0.83  &   1.15 \\
2003-01-28   &      H    &  1000    &   0.93  &   1.14 \\
2003-02-11   &      J     &   2280    &   0.93  &   1.18 \\
2003-02-13   &      Ks  &   1000    &   0.83 &   1.32 \\
2003-02-16   &      Ks  &   3500    &   1.17  &   1.25 \\                    
2003-02-16   &      H   &   1750    &   0.73  &   1.30 \\
2003-02-19   &      H   &   1750    &   0.77  &   1.48 \\       
2003-03-18   &      H   &   1750    &   0.50  &   1.44 \\       \hline
\end{tabular}
\label{ISAACdata}
\end{center}
\end{table}

The total integration times over all  nights were 3480 s ($J$), 6250 s
($H$), and 4500s ($K_s$).  For each band, observation were taken under
photometric conditions  with a seeing  often better than  1\farcs0 and
airmass below  1.5. Atmospheric conditions were average  with only the
nights of February  11th, 16th, and 19th affected by  a humidity up to
40\%. Twilight flat fields, dark frames, as well as images of standard
stars from the Persson et al.  (1998) fields, were taken daily as part
of the \isaac\  calibration plan. We downloaded the  data from the ESO
public  Science Data Archive\footnote{www.eso.org/sci/archive}  and we
reduced/calibrated them  using the updated version of  the ESO \isaac\
pipeline\footnote{http://www.eso.org/observing/dfo/quality/ISAAC/pipeline}.
For each exposure sequence,  single frames were registered and coadded
to produce a background subtracted  and cosmic-ray free image. We used
the \acs\  814W image  as a relative  reference frame to  register the
\onee\ position on the \isaac\ images.

Photometry    was    performed     using    the    {\tt    SExtractor}
software\footnote{http://www.astromatic.net/}  v2.4,  which implements
the  ``first  moment''  algorithm   \citep{kron80}.  In  view  of  the
non-optimal sky conditions, we  performed a photometric calibration of
each  image  using a  set  of  25 2MASS  stars  as  a reference.   Our
solutions turned out  to be very good, with a  r.m.s. of $\sim0.1$ mag
in  the J,  H and  Ks bands,  respectively.  The  resulting  fluxes of
source       Z      are       J=$21.53\pm0.07$,      H=$20.63\pm0.11$,
Ks=$20.53\pm0.18$.   Such  results   are  in   broad   agreement  with
\citet{wang07}  and with  \citet{fesen06}.   The upper  limits to  the
emission of \onee\ are J$\sim23.9$, H$\sim22.7$ and Ks$\sim21.7$.

\begin{table}
\begin{center}
  \caption{Results of photometry close to the position of \onee, based on \hst\ and  \vlt\  data. Values are not corrected for interstellar reddening.}
\begin{tabular}{ccc} \\ \hline
Instrument/Filter & Upper limit	& Source Z \\ \hline
\acs\ \,/F555W & 28.0 & $26.80\pm0.05$ \\
\acs\ \,/F814W & 28.1 & $24.71\pm0.01$ \\
\isaac\ \,/J   & 23.9 & $21.53\pm0.07$ \\
\isaac\ \,/H   & 22.7 & $20.63\pm0.11$ \\
\isaac\ \,/Ks  & 21.7 & $20.53\pm0.18$ \\ \hline
\end{tabular}
\label{}
\end{center}
\end{table}






\section{Discussion}
\label{disc}
Our test with \hst, based on both absolute and relative astrometry 
on multi-epoch images, firmly rules out
any physical association of Source Z with \onee.

What is source Z? Multicolor  photometry, based on the \hst\ and \vlt\
datasets, points to an unrelated background red dwarf. Flux and colors
are consistent with (although slightly redder than) an M5 star located
at $\sim5$  kpc, reddened by  E(B-V)$\sim$0.1. We note that  the \vlt\
upper  limits   based  on  our   2002  observations  in  the   R  band
\citep[R$>$27.1,][]{deluca04} are  not consistent with the source  flux as measured
with \hst\ in 2005. This could be due to some intrinsic variability of
the  dwarf  star,  and/or  to confusion effects in the \vlt\ images,
due to the  PSF wings of the  two much
brighter stars lying  a few arcsec away to  the South-West.
Of course,
in both  cases the  conclusion about the  non-association of  source Z
with \onee\ would not change.

\begin{figure}
\centering \includegraphics[width=9.0cm,clip]{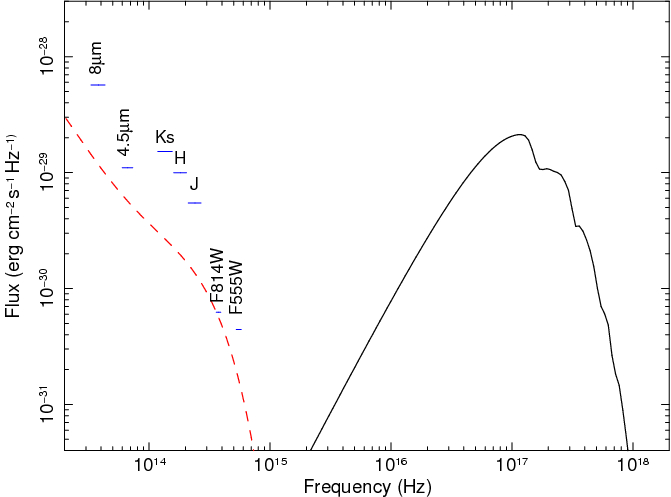}
\caption{Infrared-to-X-ray    spectral    energy   distribution    for
\onee.  Flux  limits  from  \hst\  and \vlt\  data  --  corrected  for
interstellar  reddening assuming N$_H=1.3\times10^{21}$  cm$^{-2}$ and
using the relations by  \citet{predehl95} and by \citet{cardelli89} --
are shown, together  with the limits from {\em  Spitzer} data given by
\citet{wang07}.  The  expected flux  from  a  fallback disk,  assuming
$i=60^{\circ}$,    B$=8\times10^{10}$    G,   $\dot{m}=8\times10^{11}$
g~s$^{-1}$,  $r_{out}=2\times10^{12}$ cm  (see text),  consistent with
the upper limits, is plotted as a red dashed line.  The X-ray spectrum
as observed by {\em XMM-Newton}  \citep{deluca04} is also plotted as a
solid black line. }
\label{sed}       
\end{figure}

Thus, \onee\ remains,  as yet, unidentified in the  optical/IR, as all
the         other         CCOs         observed         so         far
\citep{fesen06,deluca08b,mignani08,mignani09a,mignani09b},   with  the
only  possible   exception  of  the   source  in  the  Vela   Jr.  SNR
\citep{mignani07b}.   The upper  limits  to the  optical/IR flux  
presented here  are the
deepest available so far for a  member of the CCO class and correspond
to  an (unabsorbed) optical -- to -- X-ray   flux ratio  $F_{814W}/F_{0.3-3
keV}\,  \sim 5\times10^{-6}$.   The spectral  energy  distribution for
\onee\ is shown in Figure~\ref{sed}.

Such upper  limits virtually rule  out the possibility of  any stellar
companion tied  to the  NS in a  binary system\footnote{The  limits on
$\dot{P}$ leave some  room for a long period  system featuring a small
star.} At the distance of \onee, taking reddening into account, only a
very low-mass  star ($M\sim0.1$ M$_{\odot}$)  would be allowed.  It is
very  unlikely that  a binary  system featuring  such a  low-mass star
could survive the supernova explosion.

Combining the constraints from the  optical/IR flux limits to the ones
derived from  the timing properties of  the source \citep{gotthelf07},
we  may test  the hypothesis  of  the existence  of a  disk formed  of
supernova fallback  material, surrounding the  NS. Such a  debris disk
has  been  invoked  as  a  possible explanation  of  several  puzzling
properties  of the  X-ray source.  In  order to  compute the  expected
optical/IR flux  from such a putative  disk, we consider  a geometrically
thin,   optically  thick   fallback  disk,   locally  emitting   as  a
blackbody.   Following   a   standard   approach,  we   included   two
contributions  \citep{perna00}: (i) viscous  dissipation in
the  disk,  yielding  a  temperature  profile  $T(r)\propto  r^{-3/4}$
\citep{shakura73}; (ii) reprocessing of  X-rays emitted by the central
object. Assuming $L_X=2.2\times10^{33}$ erg s$^{-1}$ for a distance of
2.2  kpc  \citep{deluca04}  and   a  standard  disk  structure  yields
$T(r)=1100K(1-\eta)^{2/7}(R_{\odot}/r)^{3/7}$ \citep{vrtilek90}, where
$\eta$ is the X-ray albedo of  the disk. The total flux is obtained by
integrating  the emissivity  over  the disk  surface  and taking  into
account the inclination of the disk with respect to the line of sight,
as well as  the source distance.  The inner radius of  the disk is set
at  the distance  from the  star where  the magnetic  pressure  of the
rotating dipole of  the neutron star disrupts the  disk itself. Taking
into  account the  role of  viscosity in  the disk,  such a  radius is
$r_M\sim0.5R_M$   where   $R_M$  is   the   so-called  Alfven   radius
\citep[e.g.][]{frank02}.  The   outer  radius  of  the   disk  may  be
constrained  by  the  flux  limits.   We assumed  a  disk  inclination
$i=60^{\circ}$.  The  contribution   of  the  reprocessed  X-ray  flux
critically  depends on  the poorly  known value  for the  X-ray albedo
$\eta$.  Although  a value  $\eta=0.5$  has  been  assumed in  several
investigations \citep[e.g.][]{vrtilek90,perna00},  a much larger value
$\eta=0.97$ was evaluated by \citet{wang06}, based on the detection of
a disk around AXP 4U\,  0142+61. Thus, we used $\eta=0.97$ (yielding a
much lower disk luminosity) in our study. 

\begin{figure}
\centering
\includegraphics[width=9.0cm,clip]{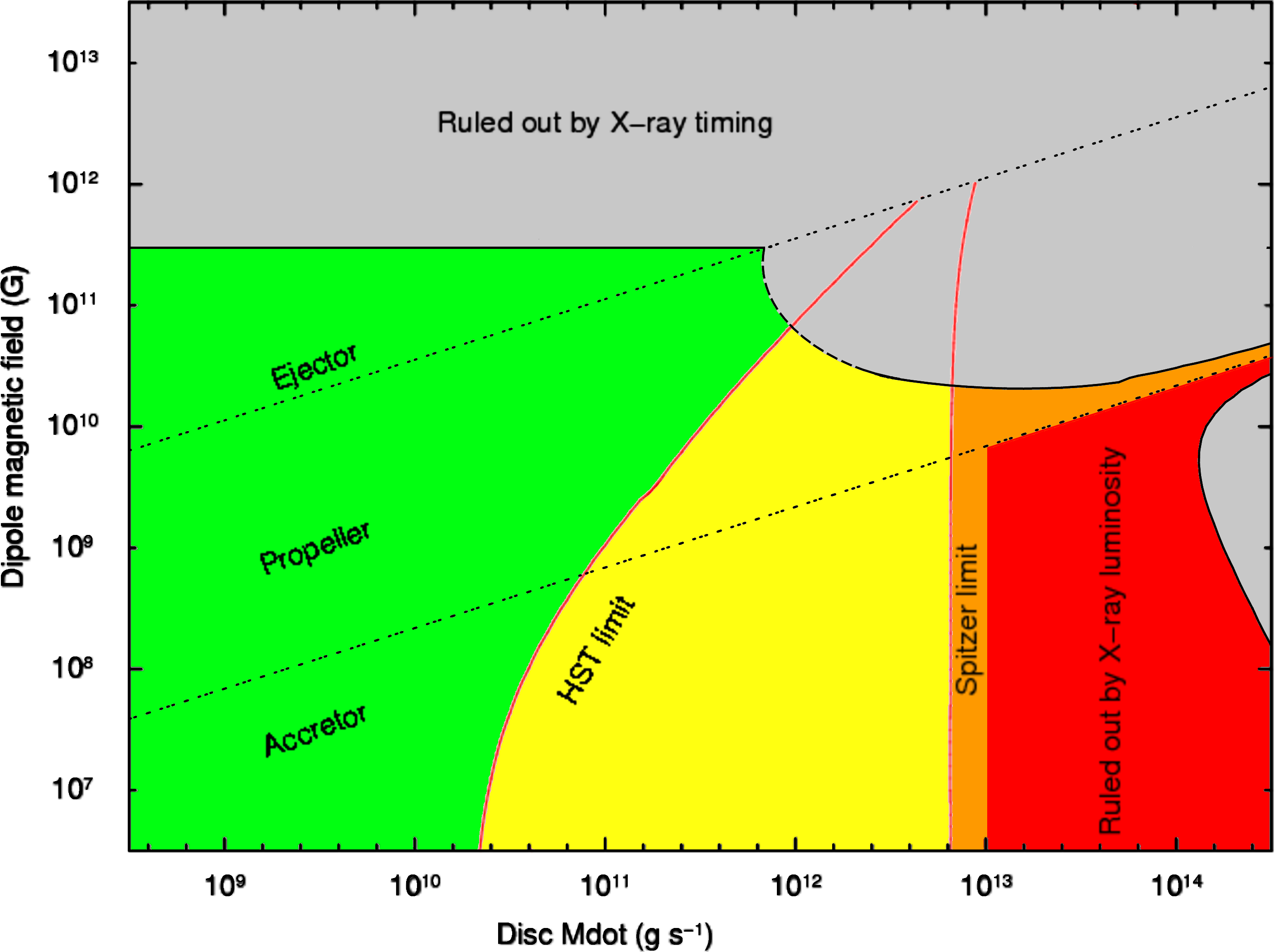}
\caption{Limits on the neutron star magnetic field and $\dot{m}$ of a putative
  fallback disk in \onee, set by X-ray timing \citep{gotthelf07} and by
optical/infrared flux limits. Colors are as follows: (grey) - ruled out by 
X-ray timing; 
(red) - ruled out by total X-ray luminosity (assuming d=2 kpc);
(orange) - consistent with X-ray constraints but already ruled out by 
former Spitzer limits \citep{wang07}; 
(yellow) - consistent with X-ray 
constraints but ruled out by our \hst\ limits; (green) - parameter region 
consistent with all constraints. }
\label{limits}       
\end{figure}

In the above assumptions, we
computed the expected flux in a  specific filter band as a function of
the NS magnetic field and disk  accretion rate $\dot{m}$, for a set of
values of the disk outer radius.  We repeated such exercise for all of
the bands  of our optical/infrared dataset. For  completeness, we also
used  the  Spitzer  $4.5\mu  m$   and  $8.4\mu  m$  bands  studied  by
\citet{wang07}. The most constraining limit  turned out to be those in
the  \hst\ F555W and F814W  bands reported  here.  In  the allowed  range of
parameters, the model  is rather insensitive to the  value of the disk
outer  radius. The  flux contributions  are almost  negligible  in all
bands for  $r_{out}$ larger than $\sim3\times10^{12}$  cm. Results are
plotted in Figure~\ref{limits} where  a red line marks the $\dot{m}$-B
field region  allowed by the combination of the F555W and F814W flux 
limits.  The region ruled out by \hst\ limits is colored in yellow.
For  comparison, we also marked (in orange) the region formerly 
ruled out by the less constraining Spitzer/IRAC $4.5\mu  m$ limit.
Flux limits in other bands are less constraining than the 
Spitzer one.

In      order       to      use      the       timing      constraints
\citep[$\dot{P}<2.5\times10^{-16}$][]{gotthelf07},  we  note that  the
interaction  of  the putative  disk  with  the  rotating neutron  star
magnetosphere  should  yield  different  regimes of  angular  momentum
transfer, according to the relative positions of $r_M$ with respect to
the neutron  star light cylinder as  well as to  the corotation radius
\citep{illarionov75}. We evaluated the expected neutron star $\dot{P}$
as a  function of  the NS  magnetic field and  disk $\dot{m}$  in such
different  regimes  (the  so-called  ejector, propeller  and  accretor
regimes)              using             standard             relations
\citep[e.g.][]{menou99,zavlin04}.   This    allows   to   identify   a
$\dot{m}$-B field  region allowed by the existing  limit on $\dot{P}$,
which is overplotted in Figure~\ref{limits}. 
The line marking the $\dot{P}$ limit has been drawn arbitrarily
(in dashed style) in the small region connecting the propeller regime
to the ejector regime (where standard relations for the
propeller torque are not valid),
to visualize the reduced efficiency of the 
propeller effect as the magnetospheric radius approaches the light 
cylinder. 
The region ruled out by X-ray timing is colored in grey.
A region in the accretor regime yielding a luminosity larger than 
$2\times10^{33}$ erg s$^{-1}$ (i.e. the total X-ray luminosity of \onee\  
assuming a distance of 2 kpc) is also ruled out. It is marked in red.

The   region   allowed  by   both   the   $\dot{P}$   limit  and   the
optical/infrared    flux    limits    is   coloured   in    green    in
Figure~\ref{limits}.
We can infer some interesting indications about the possible role of a
fallback disk.  In  the absence of any model describing a quantitative
relation between the accretion rate and the phenomenology  of the 
spectral features, we will
focus here on the issue of the  luminosity of the hot  spot. The maximum
allowed disk  $\dot{m}$ is slightly less than  $10^{12}$ g~s$^{-1}$ in
the  propeller regime. However,  as a  consequence of  the centrifugal
barrier, only a minor fraction (if  any) of such mass rate would enter
the Alfven radius. On the other hand, in the accretor regime, which is
allowed only for {\em very}  low magnetic field values (smaller than a
few  $10^{8}$ G !),  we would  expect an  accretion rate  smaller than
$\sim5\times10^{10}$ g~s$^{-1}$. In both the propeller and the accretor pictures,
the  accretion rate  on  the INS  could  yield a  luminosity of  $\sim
10^{31}$ erg  s$^{-1}$. This is  about two order of  magnitude smaller
than   the   X-ray   luminosity   of   the   hot   thermal   component
($\sim9\times10^{32}$  erg s$^{-1}$)  seen  in the  X-ray spectrum  of
\onee\ \citep{deluca04}.

Of course, such  results are model-dependent and should  be taken with
some caution.  However, they point  to an alternative  explanation for
the origin of  the high temperature and luminosity of  the hot spot on
\onee.  A very  similar picture emerged for PSR\,  J1852+0040, the CCO
at  the center  of  the Kes  79  supernova remnant.  \citet{halpern10}
succeded in measuring the $\dot{P}$  of the INS and ruled out fallback
accretion as  a heating  mechanism for the  ``hot spot'' seen  in that
source (such hot regions are ubiquitous among CCOs).  The same authors
concluded that other possible explanations of the thermal anisotropies
in   CCOs  (e.g.   localized  magnetic   heating  in   a  sunspot-like
configuration,  or anisotropic  conduction through  a  strong poloidal
magnetic field in the INS)  also suffer of several difficulties, which
also apply to the case of 1E\, 1207.4$-$5209.

It would also be interesting to constrain the total mass contained in the putative 
fallback disk,
in view of the new flux limits. However, as discussed by \citet{beckwith90},
the disk is expected to be optically thick at wavelengths shorter than $\sim1$
mm. 
Thus, a measure of the disk emission at very low frequencies would be
required to get a direct mass estimate.
While \citet{wang06}, in the case of AXP 4U\, 0142+61, could extrapolate the observed
spectral 
shape of the disk to set a flux limit in the mm band, our upper limits
do not allow us to
set constraints to the emission of the system at very low frequencies, 
hampering any direct mass estimate.
On the other hand, 
we can use the limits on the disc $\dot{m}$, coupled
to a model for fallback disk evolution as a function of time 
\citep{cannizzo90,menou99,chatterjee00,deluca06}. 
Such models expect that, after an initial (uncertain) phase during which 
supernova fallback material settles into a disk, 
the disk evolution 
is ruled by viscous processes and obeys a law $\dot{m}\propto t^{-\alpha}$,
with $\alpha$ in the range 1.18-1.25 (depending on the disk opacity).
Of course, the disc interacts with the rotating, magnetized INS and 
thus affects its spin-down history.
We assumed an age of $10^4$ yr for the system, a current disc 
$\dot{m}=5\times10^{11}$ g s$^{-1}$, a dipole
B=$8\times10^{10}$ G field for the neutron star (consistent
with a propeller regime, see Fig.\ref{limits}),
and a decay index $\alpha=1.25$. 
The model \citep[similar to the one used  by][]{deluca06}
allows to estimate the maximum disk mass 
at the onset of the disk viscous evolution ($t=t_0$, assuming the INS 
period  P$(t_0)=424$ ms) consistent 
with all of the flux and timing limits. Such an ``initial'' mass turns out to be as small 
as $10^{-6}$ M$_{\odot}$. The current disk mass (at $t=t_0+10^4$ yr), according
to the model, would be about $10^4$ times smaller. 

\section{Conclusion}

Relative and absolute astrometry on \hst\  datasets 
rules
out the  association of a faint  optical source to  \onee. Using \hst\
and \vlt\ data, we derive  very deep optical and infrared upper limits
to  any  emission  from  \onee.  Our  optical/infrared  upper  limits,
coupled to the constraints derived from the limits on $\dot{P}$, argue
against  the  possibility  that  accretion  from a  disk  of  fallback
material  could  explain the  existence  of  a  luminous, hot  thermal
component in the X-ray spectrum of this source. As for other CCOs, the
large surface thermal anisotropy  remains unexplained. 
Using published models for the evolution of fallback disks as a function of
time, we also estimated that the mass of any disk currently 
surrounding the neutron star (at an age of $\sim10^4$ yr)
should be smaller than $\sim10^{-10}$ M$_{\odot}$. 

The most likely
picture  for  \onee\ is  that  of  a young, isolated neutron  star, 
born with a rotation period very similar to the current one, 
spinning down at a tiny rate,  ruled by a dipole
field smaller  than $\sim3\times10^{11}$ G. Most  likely, the peculiar
cyclotron  features seen  in  the  X-ray spectrum  of  the source  are
generated  very  close  to the  star  surface.  A  very low  level  of
accretion ($<5\times10^{10}$  g~s$^{-1}$) cannot be  excluded, and its role  for the
production  of such features  should be  explored.  Using  the current
generation  of  ground-based or  space  observatories, more  sensitive
optical/IR  data would require  a very  large investment  of observing
time. The {\em James Webb  Space Telescope} ({\em  JWST}), to be
launched in  2014, will provide  the required sensitivity  and spatial
resolution in the near-IR to improve on our limits.



\begin{acknowledgements}

RPM  thanks STFC for support through its Rolling Grant programme and Martino Romaniello (ESO) for useful discussions. We thank the STScI for support during the re-scheduling of our \hst\ obsrevations. 

\end{acknowledgements}


\bibliographystyle{aa} 
\bibliography{1e1207.bib} 

\begin{thebibliography}{59}
\expandafter\ifx\csname natexlab\endcsname\relax\def\natexlab#1{#1}\fi

\bibitem[{{Anderson} \& {King}(1999)}]{anderson99}
{Anderson}, J. \& {King}, I.~R. 1999, \pasp, 111, 1095

\bibitem[{{Anderson} \& {King}(2003)}]{anderson03}
{Anderson}, J. \& {King}, I.~R. 2003, \pasp, 115, 113

\bibitem[{{Beckwith} {et~al.}(1990){Beckwith}, {Sargent}, {Chini}, \&
  {Guesten}}]{beckwith90}
{Beckwith}, S.~V.~W., {Sargent}, A.~I., {Chini}, R.~S., \& {Guesten}, R. 1990,
  \aj, 99, 924

\bibitem[{{Bignami} {et~al.}(2003){Bignami}, {Caraveo}, {De Luca}, \&
  {Mereghetti}}]{bignami03}
{Bignami}, G.~F., {Caraveo}, P.~A., {De Luca}, A., \& {Mereghetti}, S. 2003,
  \nat, 423, 725

\bibitem[{{Cannizzo} {et~al.}(1990){Cannizzo}, {Lee}, \&
  {Goodman}}]{cannizzo90}
{Cannizzo}, J.~K., {Lee}, H.~M., \& {Goodman}, J. 1990, \apj, 351, 38

\bibitem[{{Caraveo} {et~al.}(2001){Caraveo}, {De Luca}, {Mignani}, \&
  {Bignami}}]{caraveo01}
{Caraveo}, P.~A., {De Luca}, A., {Mignani}, R.~P., \& {Bignami}, G.~F. 2001,
  \apj, 561, 930

\bibitem[{{Cardelli} {et~al.}(1989){Cardelli}, {Clayton}, \&
  {Mathis}}]{cardelli89}
{Cardelli}, J.~A., {Clayton}, G.~C., \& {Mathis}, J.~S. 1989, \apj, 345, 245

\bibitem[{{Chatterjee} {et~al.}(2000){Chatterjee}, {Hernquist}, \&
  {Narayan}}]{chatterjee00}
{Chatterjee}, P., {Hernquist}, L., \& {Narayan}, R. 2000, \apj, 534, 373

\bibitem[{{Clampin} {et~al.}(2000){Clampin}, {Ford}, {Bartko}, {Bely},
  {Broadhurst}, {Burrows}, {Cheng}, {Crocker}, {Franx}, {Feldman},
  {Golimowski}, {Hartig}, {Illingworth}, {Kimble}, {Lesser}, {Miley},
  {Postman}, {Rafal}, {Rosati}, {Sparks}, {Tsvetanov}, {White}, {Sullivan},
  {Volmer}, \& {LaJeunesse}}]{clampin00}
{Clampin}, M., {Ford}, H.~C., {Bartko}, F., {et~al.} 2000, in Presented at the
  Society of Photo-Optical Instrumentation Engineers (SPIE) Conference, Vol.
  4013, Society of Photo-Optical Instrumentation Engineers (SPIE) Conference
  Series, ed. {J.~B.~Breckinridge \& P.~Jakobsen}, 344--351

\bibitem[{{de Luca}(2008)}]{deluca08a}
{de Luca}, A. 2008, in American Institute of Physics Conference Series, Vol.
  983, 40 Years of Pulsars: Millisecond Pulsars, Magnetars and More, ed.
  {C.~Bassa, Z.~Wang, A.~Cumming, \& V.~M.~Kaspi}, 311--319

\bibitem[{{De Luca} {et~al.}(2009){De Luca}, {Caraveo}, {Esposito}, \&
  {Hurley}}]{deluca09}
{De Luca}, A., {Caraveo}, P.~A., {Esposito}, P., \& {Hurley}, K. 2009, \apj,
  692, 158

\bibitem[{{De Luca} {et~al.}(2006){De Luca}, {Caraveo}, {Mereghetti}, {Tiengo},
  \& {Bignami}}]{deluca06}
{De Luca}, A., {Caraveo}, P.~A., {Mereghetti}, S., {Tiengo}, A., \& {Bignami},
  G.~F. 2006, Science, 313, 814

\bibitem[{{De Luca} {et~al.}(2004){De Luca}, {Mereghetti}, {Caraveo}, {Moroni},
  {Mignani}, \& {Bignami}}]{deluca04}
{De Luca}, A., {Mereghetti}, S., {Caraveo}, P.~A., {et~al.} 2004, \aap, 418,
  625

\bibitem[{{De Luca} {et~al.}(2000){De Luca}, {Mignani}, \&
  {Caraveo}}]{deluca99}
{De Luca}, A., {Mignani}, R.~P., \& {Caraveo}, P.~A. 2000, \aap, 354, 1011

\bibitem[{{De Luca} {et~al.}(2007){De Luca}, {Mignani}, {Caraveo}, \&
  {Bignami}}]{deluca07}
{De Luca}, A., {Mignani}, R.~P., {Caraveo}, P.~A., \& {Bignami}, G.~F. 2007,
  \apjl, 667, L77

\bibitem[{{De Luca} {et~al.}(2008){De Luca}, {Mignani}, {Zaggia}, {Beccari},
  {Mereghetti}, {Caraveo}, \& {Bignami}}]{deluca08b}
{De Luca}, A., {Mignani}, R.~P., {Zaggia}, S., {et~al.} 2008, \apj, 682, 1185

\bibitem[{{Fesen} {et~al.}(2006){Fesen}, {Pavlov}, \& {Sanwal}}]{fesen06}
{Fesen}, R.~A., {Pavlov}, G.~G., \& {Sanwal}, D. 2006, \apj, 636, 848

\bibitem[{{Frank} {et~al.}(2002){Frank}, {King}, \& {Raine}}]{frank02}
{Frank}, J., {King}, A., \& {Raine}, D.~J. 2002, {Accretion Power in
  Astrophysics: Third Edition}, ed. {Frank, J., King, A., \& Raine, D.~J.}

\bibitem[{{Giacani} {et~al.}(2000){Giacani}, {Dubner}, {Green}, {Goss}, \&
  {Gaensler}}]{giacani00}
{Giacani}, E.~B., {Dubner}, G.~M., {Green}, A.~J., {Goss}, W.~M., \&
  {Gaensler}, B.~M. 2000, \aj, 119, 281

\bibitem[{{Gotthelf} \& {Halpern}(2007)}]{gotthelf07}
{Gotthelf}, E.~V. \& {Halpern}, J.~P. 2007, \apjl, 664, L35

\bibitem[{{Gotthelf} \& {Halpern}(2009)}]{gotthelf09}
{Gotthelf}, E.~V. \& {Halpern}, J.~P. 2009, \apjl, 695, L35

\bibitem[{{Halpern} \& {Gotthelf}(2010)}]{halpern10}
{Halpern}, J.~P. \& {Gotthelf}, E.~V. 2010, \apj, 709, 436

\bibitem[{{Halpern} {et~al.}(2007){Halpern}, {Gotthelf}, {Camilo}, \&
  {Seward}}]{halpern07}
{Halpern}, J.~P., {Gotthelf}, E.~V., {Camilo}, F., \& {Seward}, F.~D. 2007,
  \apj, 665, 1304

\bibitem[{{Hobbs} {et~al.}(2005){Hobbs}, {Lorimer}, {Lyne}, \&
  {Kramer}}]{hobbs05}
{Hobbs}, G., {Lorimer}, D.~R., {Lyne}, A.~G., \& {Kramer}, M. 2005, \mnras,
  360, 974

\bibitem[{{Illarionov} \& {Sunyaev}(1975)}]{illarionov75}
{Illarionov}, A.~F. \& {Sunyaev}, R.~A. 1975, \aap, 39, 185

\bibitem[{{Kron}(1980)}]{kron80}
{Kron}, R.~G. 1980, \apjs, 43, 305

\bibitem[{{Lasker} {et~al.}(2008){Lasker}, {Lattanzi}, {McLean}, {Bucciarelli},
  {Drimmel}, {Garcia}, {Greene}, {Guglielmetti}, {Hanley}, {Hawkins},
  {Laidler}, {Loomis}, {Meakes}, {Mignani}, {Morbidelli}, {Morrison},
  {Pannunzio}, {Rosenberg}, {Sarasso}, {Smart}, {Spagna}, {Sturch},
  {Volpicelli}, {White}, {Wolfe}, \& {Zacchei}}]{lasker08}
{Lasker}, B.~M., {Lattanzi}, M.~G., {McLean}, B.~J., {et~al.} 2008, \aj, 136,
  735

\bibitem[{{Lattanzi} {et~al.}(1997){Lattanzi}, {Capetti}, \&
  {Macchetto}}]{lattanzi97}
{Lattanzi}, M.~G., {Capetti}, A., \& {Macchetto}, F.~D. 1997, \aap, 318, 997

\bibitem[{{Liu} {et~al.}(2006){Liu}, {Yuan}, {Chen}, \& {You}}]{liu06}
{Liu}, D.~B., {Yuan}, A.~F., {Chen}, L., \& {You}, J.~H. 2006, \apj, 644, 439

\bibitem[{{Menou} {et~al.}(1999){Menou}, {Esin}, {Narayan}, {Garcia}, {Lasota},
  \& {McClintock}}]{menou99}
{Menou}, K., {Esin}, A.~A., {Narayan}, R., {et~al.} 1999, \apj, 520, 276

\bibitem[{{Mereghetti} {et~al.}(2002){Mereghetti}, {De Luca}, {Caraveo},
  {Becker}, {Mignani}, \& {Bignami}}]{mereghetti02}
{Mereghetti}, S., {De Luca}, A., {Caraveo}, P.~A., {et~al.} 2002, \apj, 581,
  1280

\bibitem[{{Mignani} {et~al.}(2007{\natexlab{a}}){Mignani}, {Bagnulo}, {de
  Luca}, {Israel}, {Lo Curto}, {Motch}, {Perna}, {Rea}, {Turolla}, \&
  {Zane}}]{mignani07a}
{Mignani}, R.~P., {Bagnulo}, S., {de Luca}, A., {et~al.} 2007{\natexlab{a}},
  \apss, 308, 203

\bibitem[{{Mignani} {et~al.}(2002){Mignani}, {De Luca}, {Caraveo}, \&
  {Becker}}]{mignani02}
{Mignani}, R.~P., {De Luca}, A., {Caraveo}, P.~A., \& {Becker}, W. 2002, \apjl,
  580, L147

\bibitem[{{Mignani} {et~al.}(2009{\natexlab{a}}){Mignani}, {de Luca},
  {Mereghetti}, \& {Caraveo}}]{mignani09a}
{Mignani}, R.~P., {de Luca}, A., {Mereghetti}, S., \& {Caraveo}, P.~A.
  2009{\natexlab{a}}, \aap, 500, 1211

\bibitem[{{Mignani} {et~al.}(2009{\natexlab{b}}){Mignani}, {de Luca}, \&
  {Pellizzoni}}]{mignani09b}
{Mignani}, R.~P., {de Luca}, A., \& {Pellizzoni}, A. 2009{\natexlab{b}}, \aap,
  508, 779

\bibitem[{{Mignani} {et~al.}(2007{\natexlab{b}}){Mignani}, {de Luca}, {Zaggia},
  {Sester}, {Pellizzoni}, {Mereghetti}, \& {Caraveo}}]{mignani07b}
{Mignani}, R.~P., {de Luca}, A., {Zaggia}, S., {et~al.} 2007{\natexlab{b}},
  \aap, 473, 883

\bibitem[{{Mignani} {et~al.}(2010){Mignani}, {Sartori}, {De Luca}, {Rudack},
  {Slowikowska}, {Kanbach}, \& {Caraveo}}]{mignani10}
{Mignani}, R.~P., {Sartori}, A., {De Luca}, A., {et~al.} 2010, ArXiv e-prints

\bibitem[{{Mignani} {et~al.}(2008){Mignani}, {Zaggia}, {de Luca}, {Perna},
  {Bassan}, \& {Caraveo}}]{mignani08}
{Mignani}, R.~P., {Zaggia}, S., {de Luca}, A., {et~al.} 2008, \aap, 484, 457

\bibitem[{{Monet}(1998)}]{monet98}
{Monet}, D. 1998, {USNO-A2.0}, ed. {Monet, D.}

\bibitem[{{Monet} {et~al.}(2003){Monet}, {Levine}, {Canzian}, {Ables}, {Bird},
  {Dahn}, {Guetter}, {Harris}, {Henden}, {Leggett}, {Levison}, {Luginbuhl},
  {Martini}, {Monet}, {Munn}, {Pier}, {Rhodes}, {Riepe}, {Sell}, {Stone},
  {Vrba}, {Walker}, {Westerhout}, {Brucato}, {Reid}, {Schoening}, {Hartley},
  {Read}, \& {Tritton}}]{monet03}
{Monet}, D.~G., {Levine}, S.~E., {Canzian}, B., {et~al.} 2003, \aj, 125, 984

\bibitem[{{Pavlov} {et~al.}(2002){Pavlov}, {Sanwal}, {Garmire}, \&
  {Zavlin}}]{pavlov02}
{Pavlov}, G.~G., {Sanwal}, D., {Garmire}, G.~P., \& {Zavlin}, V.~E. 2002, in
  Astronomical Society of the Pacific Conference Series, Vol. 271, Neutron
  Stars in Supernova Remnants, ed. {P.~O.~Slane \& B.~M.~Gaensler}, 247--+

\bibitem[{{Pavlov} {et~al.}(2004){Pavlov}, {Sanwal}, \& {Teter}}]{pavlov04}
{Pavlov}, G.~G., {Sanwal}, D., \& {Teter}, M.~A. 2004, in IAU Symposium, Vol.
  218, Young Neutron Stars and Their Environments, ed. {F.~Camilo \&
  B.~M.~Gaensler}, 239--+

\bibitem[{{Perna} {et~al.}(2000){Perna}, {Hernquist}, \& {Narayan}}]{perna00}
{Perna}, R., {Hernquist}, L., \& {Narayan}, R. 2000, \apj, 541, 344

\bibitem[{{Potekhin}(2010)}]{potekhin10}
{Potekhin}, A.~Y. 2010, ArXiv e-prints

\bibitem[{{Predehl} \& {Schmitt}(1995)}]{predehl95}
{Predehl}, P. \& {Schmitt}, J.~H.~M.~M. 1995, \aap, 293, 889

\bibitem[{{Roger} {et~al.}(1988){Roger}, {Milne}, {Kesteven}, {Wellington}, \&
  {Haynes}}]{roger88}
{Roger}, R.~S., {Milne}, D.~K., {Kesteven}, M.~J., {Wellington}, K.~J., \&
  {Haynes}, R.~F. 1988, \apj, 332, 940

\bibitem[{{Sanwal} {et~al.}(2002){Sanwal}, {Pavlov}, {Zavlin}, \&
  {Teter}}]{sanwal02}
{Sanwal}, D., {Pavlov}, G.~G., {Zavlin}, V.~E., \& {Teter}, M.~A. 2002, \apjl,
  574, L61

\bibitem[{{Shakura} \& {Sunyaev}(1973)}]{shakura73}
{Shakura}, N.~I. \& {Sunyaev}, R.~A. 1973, \aap, 24, 337

\bibitem[{{Sirianni} {et~al.}(2005){Sirianni}, {Jee}, {Ben{\'{\i}}tez},
  {Blakeslee}, {Martel}, {Meurer}, {Clampin}, {De Marchi}, {Ford}, {Gilliland},
  {Hartig}, {Illingworth}, {Mack}, \& {McCann}}]{sirianni05}
{Sirianni}, M., {Jee}, M.~J., {Ben{\'{\i}}tez}, N., {et~al.} 2005, \pasp, 117,
  1049

\bibitem[{{Skrutskie} {et~al.}(2006){Skrutskie}, {Cutri}, {Stiening},
  {Weinberg}, {Schneider}, {Carpenter}, {Beichman}, {Capps}, {Chester},
  {Elias}, {Huchra}, {Liebert}, {Lonsdale}, {Monet}, {Price}, {Seitzer},
  {Jarrett}, {Kirkpatrick}, {Gizis}, {Howard}, {Evans}, {Fowler}, {Fullmer},
  {Hurt}, {Light}, {Kopan}, {Marsh}, {McCallon}, {Tam}, {Van Dyk}, \&
  {Wheelock}}]{skrutskie06}
{Skrutskie}, M.~F., {Cutri}, R.~M., {Stiening}, R., {et~al.} 2006, \aj, 131,
  1163

\bibitem[{{Storey} {et~al.}(1992){Storey}, {Staveley-Smith}, {Manchester}, \&
  {Kesteven}}]{storey92}
{Storey}, M.~C., {Staveley-Smith}, L., {Manchester}, R.~N., \& {Kesteven},
  M.~J. 1992, \aap, 265, 752

\bibitem[{{Suleimanov} {et~al.}(2010){Suleimanov}, {Pavlov}, \&
  {Werner}}]{suleimanov10}
{Suleimanov}, V.~F., {Pavlov}, G.~G., \& {Werner}, K. 2010, ArXiv e-prints

\bibitem[{{Tuohy} \& {Garmire}(1980)}]{tuohy80}
{Tuohy}, I. \& {Garmire}, G. 1980, \apjl, 239, L107

\bibitem[{{Vrtilek} {et~al.}(1990){Vrtilek}, {Raymond}, {Garcia}, {Verbunt},
  {Hasinger}, \& {Kurster}}]{vrtilek90}
{Vrtilek}, S.~D., {Raymond}, J.~C., {Garcia}, M.~R., {et~al.} 1990, \aap, 235,
  162

\bibitem[{{Wang} {et~al.}(2006){Wang}, {Chakrabarty}, \& {Kaplan}}]{wang06}
{Wang}, Z., {Chakrabarty}, D., \& {Kaplan}, D.~L. 2006, \nat, 440, 772

\bibitem[{{Wang} {et~al.}(2007){Wang}, {Kaplan}, \& {Chakrabarty}}]{wang07}
{Wang}, Z., {Kaplan}, D.~L., \& {Chakrabarty}, D. 2007, \apj, 655, 261

\bibitem[{{Woods} {et~al.}(2007){Woods}, {Zavlin}, \& {Pavlov}}]{woods07}
{Woods}, P.~M., {Zavlin}, V.~E., \& {Pavlov}, G.~G. 2007, \apss, 308, 239

\bibitem[{{Zavlin} {et~al.}(2004){Zavlin}, {Pavlov}, \& {Sanwal}}]{zavlin04}
{Zavlin}, V.~E., {Pavlov}, G.~G., \& {Sanwal}, D. 2004, \apj, 606, 444

\bibitem[{{Zavlin} {et~al.}(2000){Zavlin}, {Pavlov}, {Sanwal}, \&
  {Tr{\"u}mper}}]{zavlin00}
{Zavlin}, V.~E., {Pavlov}, G.~G., {Sanwal}, D., \& {Tr{\"u}mper}, J. 2000,
  \apjl, 540, L25

\end{thebibliography}

%
%

\end{document}